\documentclass[aps,preprint,epsfig,rotate]{revtex4}
\usepackage{graphicx}
\usepackage{bm}
\usepackage{epsfig}

\begin{document}
\title{Uehling potential and lowest-order corrections on vacuum polarization to the cross sections of some QED processes}  

\author{Alexei M. Frolov}
 \email[E--mail address: ]{alex1975frol@gmail.com} 


\affiliation{Department of Applied Mathematics \\
 University of Western Ontario, London, Ontario N6H 5B7, Canada}

\date{\today}

\begin{abstract}

Properties and different representations of the Uehling potential are investigated. Based on these properties and by using our formulas 
for the Fourier transform of the Uehling potential we have developed the new analytical, logically closed and physically transparent 
procedure which can be used to evaluate the lowest-order vacuum polarization correction to the cross sections of a number of QED processes, 
including the Mott electron scattering, bremsstrahlung, creation and/or annihilation of the $(e^{-}, e^{+})-$pair in the field of a heavy 
Coulomb center, e.g., atomic nucleus. \\

\noindent 
PACS number(s): 12.20.-m , 12.20.Ds and 31.30.Jv \\

\noindent 

\noindent 
This version of the manuscriot is quite close to the final version published in European Physical Journal A - Hadrons and Nuclei, {\bf 57}, 79 
(2021) (DOI: doi.org/10.1140/epja/s10050-021-00394-y). 

\end{abstract}
\maketitle

\section{Introduction}

In this short communication we investigate radiative corrections to the Coulomb law. As is well known (see, e.g., \cite{Grein}, \cite{AB}, 
\cite{BLP}) these corrections can explicitly be described as a result of vacuum polarization around a point electric charge. In the 
lowest-order approximation the vacuum polarization, which always arises around an arbitrary (point) electrical charge, is described by the 
Uehling potential \cite{Uehl} (see, also, \cite{Grein} and \cite{AB}). For atomic and Coulomb few- and many-body systems the Uehling 
potential \cite{Uehl} generates a small correction on vacuum polarization which must be added to the leading contribution from the original 
Coulomb potential. The sum of the original Coulomb and Uehling potentials is represent in the following integral form \cite{Grein} 
\begin{eqnarray}    
  V(r) = - \frac{Q e}{r} + U(r) = - \frac{Q e}{r} \Bigl[ 1 + \frac{2 \alpha}{3 \pi} \int_{1}^{\infty} d\varsigma \Bigl( 1 + \frac{1}{2 
 \varsigma^{2}} \Bigr) \frac{\sqrt{\varsigma^{2} - 1}}{\varsigma^{2}} \exp(- 2 m_e r \varsigma) \Bigr] \; \; \label{first} 
\end{eqnarray}
where $- Q e$ is the electric charge of the central atomic nucleus, $e$ is the charge of the electron $e$, where $e < 0$, $\alpha = 
\frac{e^{2}}{\hbar c} \approx \frac{1}{137}$ is the fine-structure constant, while $m_e$ is the electron mass at rest and $r$ is the 
electron-nuclear distance. In this formula and everywhere below in this study we apply the relativistic units, where $\hbar = 1$ and 
$c = 1$ \cite{Mandl}. In these units the Uehling potential $U(r)$ from Eq.(\ref{first}) is written in the form
\begin{eqnarray}    
 U(r) = - \frac{2 Q e \alpha}{3 \pi r} \int_{1}^{\infty} \Bigl( 1 + \frac{1}{2 \varsigma^{2}} \Bigr) \frac{\sqrt{\varsigma^{2} - 
 1}}{\varsigma^{2}} \exp(- 2 m_e r \varsigma) d\varsigma \; \; \label{U1} 
\end{eqnarray} 
Below we shall assume that such a vacuum polarization is produced and described by the Uehling potential, Eq.(\ref{U1}), only. In other words, in 
the course of our analysis all higher-order corrections for vacuum polarization will be ignored. Applications of the Uehling potential to various
atomic and muon-atomic systems can be found, e.g., in \cite{Dubl}, \cite{Plum} and \cite{Fro3} and referenes mentioned in these papers. Here we 
will not discuss such applications, since they are quite different from the goals of this work. 

The main goal of this study is to analyze the basic properties of Uehling potential, Eq.(\ref{U1}). We also derive a few different representations 
of the Uehling potential, Eq.(\ref{U1}). Then, by using our formula(s) for the Fourier transform of the Uehling potential we have developed the new 
transparent and covariant procedure which allows one to evaluate the lowest-order correction on vacuum polarization to the cross sections of a 
number of QED processes. As follows from the results of our study, the Uehling potential, Eq.(\ref{U1}), can be included in the set of fundamental 
and covariant QED rules which is used for analytical calculations of the $S_{fi}$-matrix elements from the very beginning.  

\section{Analytical formulas for the Uehling potential}

In numerous books and textbooks on Quantum Electrodynamics (see, e.g., \cite{Grein}, \cite{AB}, \cite{BLP} and references therein) one usually finds 
the following statement (or one of its variations): the `closed analytical expression for the Uehling potential, which is correct for arbitrary 
electron-nuclear distances $r$, does not exist'. A different form of the same statement is: `analytical formula for the Uehling potential can be found 
only for very small $r$ and/or for very large $r$' (see, e.g., \cite{BLP}). For the first time this statement appeared in some QED book(s) published 
in early 1950's. In fact, quite a few of such books were published immediately after appearence of the fundamental papers by Feynman \cite{Feyn} and 
Dyson \cite{Dyson}. Since then this statement has not been verified, and now it simply wanders from one QED-book into another. In this Section we want 
to show that this statement is wrong and misleading. Very likely, that whoever wrote and published for the first time was not very familiar with the 
theory of Bessel functions. 

Indeed, let us derive a few simple and explicit analytical formulas for the Uehling potential $U(r)$ from Eq.(\ref{U1}). First, we introduce the 
new variable $\varsigma = \cosh \chi$ in Eq.(\ref{U1}), where $0 \le \chi < +\infty$, and obtain the following expression 
\begin{eqnarray}    
 U(r) &=& - \frac{2 Q e \alpha}{3 \pi r} \int_{0}^{\infty} \Bigl( 1 - \frac{1}{\cosh^{2}\chi} \Bigr) \Bigl( 1 + \frac{1}{2 \cosh^{2}\chi} \Bigr) 
 \exp(- 2 m_e r \cosh\chi) d\chi \nonumber \\ 
 &=& - \frac{2 Q e \alpha}{3 \pi r} \int_{0}^{\infty} \Bigl( 1 - \frac{1}{2 \cosh^{2}\chi} - \frac{1}{2 \cosh^{4}\chi} \Bigr) 
 \exp(- 2 m_e r \cosh\chi) d\chi \nonumber \\ 
 &=& - \frac{2 Q e \alpha}{3 \pi r} \Bigl[ \int_{0}^{\infty} \exp(- 2 m_e r \cosh\chi) d\chi - \frac12 \int_{0}^{\infty} \frac{\exp(- 2 m_e r 
 \cosh\chi) d\chi}{\cosh^{2}\chi} \nonumber \\
 & & - \frac12 \int_{0}^{\infty} \frac{\exp(- 2 m_e r \cosh\chi) d\chi}{\cosh^{4}\chi} \Bigr] \; \; \label{U1a} 
\end{eqnarray} 
where the first integral is the modified Bessel function $K_{0}(2 m_e r)$ of the second kind, which is also called the Macdonald function and/or 
Hankel function of purely imaginary argument \cite{AS}, \cite{GR} and \cite{Wats}, while the second and third integrals in Eq.(\ref{U1a}) are the 
multiple integrals of the $K_{0}(2 m_e r)$ function. Briefly, we can write (see, e.g., Chapter 11 in \cite{AS})  
\begin{eqnarray} 
 K_{0}(2 m_e r) \equiv Ki_{0}(2 m_e r) = \int_{0}^{\infty} \exp(- 2 m_e r \cosh\chi) d\chi \; \; \label{K0}
\end{eqnarray}
and
\begin{eqnarray}
 Ki_{n}(2 m_e r) = \int_{2 m_e r}^{+\infty} Ki_{n-1}(y) dy = \int_{0}^{\infty} \frac{\exp(- 2 m_e r \cosh\chi) d\chi}{\cosh^{n}\chi} \; \; , 
 \; \; \label{Ki}
\end{eqnarray} 
for $n = 1, 2, 3, \ldots$ (here $n$ is a non-negative integer number). The explicit formula for the $K_{0}(z)$ function is
\begin{eqnarray}
  K_{0}(z) = \sum^{\infty}_{k=0} \Bigl[ \psi(k + 1) - \ln z + \ln 2 \Bigr] \frac{z^{2 k}}{2^{2 k} (k!)^2} = \sum^{\infty}_{k=0} \psi(k + 1) 
  \frac{z^{2 k}}{2^{2 k} (k!)^2} - \ln\Bigl(\frac{z}{2}\Bigr) \sum^{\infty}_{k=0} \frac{z^{2 k}}{2^{2 k} (k!)^2} \; \; \label{K0Frm}
\end{eqnarray} 
where $\psi(x) = \frac{d \Gamma(x)}{dx}$ is the $\psi-$function, while $\Gamma(x)$ is the gamma-function \cite{GR} (or Euler's integral of the 
second type). There are many formulas which represent this $\psi-$function in different forms and some of these forms are very convenient for 
numerical calculations. For instance, the following series representation 
\begin{eqnarray}
  \psi(x) = \ln x - \sum^{\infty}_{n=0} \Bigl[ \frac{1}{x + n} - \ln\Bigl( 1 + \frac{1}{x + n}\Bigr)\Bigr] \; \; \label{psi}
\end{eqnarray}  
is widely used to determine numerical values of the $\psi(x)$ function. 

Based on Eqs.(\ref{U1a}) - (\ref{Ki}) one can write the following explicit formula for the Uehling potential 
\begin{eqnarray}    
 U(r) = - \frac{Q e \alpha}{3 \pi r} \Bigl[ 2 K_{0}(2 m_e r) - Ki_{2}(2 m_e r) - Ki_{4}(2 m_e r) \Bigr] \; \; \label{form1}  
\end{eqnarray}   
This is the first analytical formula for the Uehling potential \cite{Fro1}. Advantage of this simple formula is obvious, since all coefficients
in this formula are the numerical constants. However, by using the well known recursive relation \cite{AS} for the $Ki_{n}(z)$ functions (or 
integrals) 
\begin{eqnarray} 
  n Ki_{n+1}(z) = -z Ki_{n}(z) + (n - 1) Ki_{n-1}(z) + z Ki_{n-2}(z) \; \; \label{recurs}
\end{eqnarray}
for $n$ = 2, 3, 4, $\ldots$, one can reduce the formula, Eq.(\ref{form1}), to a different form 
\begin{eqnarray}    
 U(r) &=& -\frac{Q e \alpha}{18 \pi r} \Bigl[ (12 + z^{2}) Ki_{0}(z) - z Ki_{1}(z) - (10 + z^{2}) Ki_{2}(z)\Bigr] \; \; \nonumber \\
 &=& - \frac{Q e \alpha}{18 \pi r} \Bigl[ (1 + q(z)) K_{0}(z) - z Ki_{1}(z) + (1 - q(z)) Ki_{2}(z) \Bigr] \; \; \label{formula1} 
\end{eqnarray}  
where $z = 2 m_e r$ and $q(z) = 11 + z^{2}$ is the quadratic function of $z$. This is the second analytical formula known \cite{Fro1} for the 
Uehling potential. This formula contains only the two lowest multiple integrals of the $K_{0}(2 m_e r)$ function, but coefficients in this 
formula are the two quadratic functions of $z$ and one linear function of $z = 2 m_e r$. By using more formulas known for the modified Bessel 
functions one can derive more analytical formulas for the Uehling potential $U(r)$.  

\section{Fourier transform of the Uehling potential}

Let us obtain the explicit formulas for the Fourier transform of the Uehling potential \cite{QEDT}. This problem was considered earlier in 
\cite{Fro2} and \cite{Pauli}. First, we derive the explicit formula for the three-dimensional Fourier transform \cite{QEDT} of an arbitrary 
Yukawa-type potential ($\simeq \frac{\exp(- 2 m_e \varsigma)}{r}$) 
\begin{eqnarray}  
 & & \int_{0}^{+\infty} \int_{0}^{+\infty} \int_{0}^{+\infty} d^{3}{\bf x} \exp(\imath {\bf q} {\bf x}) \frac{\exp(- 2 m_e \varsigma)}{r} = 
 \frac{4 \pi}{q} \int_{0}^{\infty} dr \sin(q r) \exp(-2 m_e \varsigma r) \; \nonumber \\
 &=& \frac{\pi}{m^{2}_e [\varsigma^{2} + b^{2}]} \; \; \label{Four1}
\end{eqnarray}  
where $b = \frac{q}{2 m_e}$ (or $b^{2} = \frac{q^{2}}{4 m^{2}_e}$). Here we applied the formula (5) in Eq.(\ref{3.944}) form \cite{GR} which is 
\begin{eqnarray}  
 \int_{0}^{+\infty} x^{\mu - 1} \exp(-\beta x) \sin(\delta x) dx = \frac{\Gamma(\mu)}{\Bigl(\beta^{2} + \delta^{2}\Bigr)^{\frac{\mu}{2}}} 
 \sin\Bigl(\mu \arctan \frac{\delta}{\beta}\Bigr) \; \; \label{3.944}
\end{eqnarray} 
For $\mu = 1$ one finds from Eq.(\ref{3.944}) 
\begin{eqnarray}  
 \int_{0}^{+\infty} \exp(-\beta x) \sin(\delta x) dx = \frac{\delta}{\beta^{2} + \delta^{2}} \; \;  \label{3.945}
\end{eqnarray} 

By using the formula, Eq.(\ref{Four1}), we reduce the problem of analytical computation of the Fourier transform of the Uehling potential 
$\tilde{U}(q)$ to the following one-dimensional integral 
\begin{eqnarray}
 \tilde{U}(q) = - \frac{2 Q e \alpha}{3 m^{2}_{e}} \int_{1}^{+\infty} dt \Bigl(1 + \frac{1}{2 t^{2}}\Bigr) \frac{\sqrt{t^{2} - 1}}{t^{2} 
 (t^{2} + b^{2})} \; \; \label{Uq}
\end{eqnarray} 
where $b^{2} = \frac{q^{2}}{4 m^{2}_e}$. The explixit formula for this integral is derived by introducing the new variable $v = 
\frac{\sqrt{t^{2} - 1}}{t}$. Now, the integral from Eq.(\ref{Uq}) is reduced to the following form
\begin{eqnarray}
 \int_{1}^{+\infty} dt \Bigl(1 + \frac{1}{2 t^{2}}\Bigr) \frac{\sqrt{t^{2} - 1}}{t^{2} (t^{2} + b^{2})} = \frac{1}{2 b^{2}} \int_{0}^{1} dv 
 \frac{(3 - v^{2}) v^{2}}{d^{2} - v^{2}} \; \; \label{FUq}
\end{eqnarray} 
where $d^{2} = \frac{b^{2} + 1}{b^{2}} = 1 + \frac{1}{b^{2}} = \frac{q^{2} + 4 m^{2}}{4 m^{2}}$. After a few additional steps we obtain the 
following formula for the Fourier transform of the Uehling potential 
\begin{eqnarray}
 \tilde{U}(q) = - \frac{4 Q e \alpha}{3 q^{2}} \Bigl\{ - \frac53 + \frac{4 m^{2}_e}{q^{2}} + \Bigl[1 - \frac12 
 \Bigl(\frac{4 m^{2}_e}{q^{2}}\Bigr)\Bigr] \ln\Bigl(\frac{\sqrt{q^{2} + 4 m^{2}_e} + 2 m_e}{\sqrt{q^{2} + 4 m^{2}_e} - 
 2 m_e}\Bigr) \Bigr\} \; \; \label{FUqfinal} 
\end{eqnarray} 
which exactly coincides with the result obtained in \cite{Fro2} (see also \cite{Pauli}). The sum of this Fourier transform of the Uehling 
potential with the original Coulomb potenial $W(q)$ is 
\begin{eqnarray}
 & & W(q) = - \frac{4 \pi Q e}{q^{2}} + \tilde{U}(q) \nonumber \\
 &=& -\frac{4 \pi Q e}{q^{2}} \Biggl\{ 1 + \frac{\alpha}{3 \pi} \Bigl\{ -\frac53 + 
 \frac{4 m^{2}_e}{q^{2}} + \Bigl[1 - \frac12 \Bigl(\frac{4 m^{2}_e}{q^{2}}\Bigr)\Bigr] \sqrt{1 + \frac{4 m^{2}_e}{q^{2}}} \ln\Bigl(\frac{\sqrt{q^{2} 
 + 4 m^{2}_e} + 2 m_e}{\sqrt{q^{2} + 4 m^{2}_e} - 2 m_e}\Bigr) \Bigr\} \Biggr\} \; \; \nonumber \\
 &=& -\frac{4 \pi Q e}{q^{2}} \Biggl\{ 1 + \frac{\alpha}{3 \pi} \Bigl[ -\frac53 + \frac{1}{b^{2}} - \Bigl(1 - \frac{1}{2 b^{2}}\Bigr) 
 \sqrt{1 + \frac{1}{b^{2}}} \ln\Bigl(\frac{\sqrt{b^{2} + 1} + 1}{\sqrt{b^{2} + 1} - 1}\Bigr) \Bigr] \Biggr\} \; \; \label{finalF} 
\end{eqnarray} 
where $b^{2} = \frac{q^{2}}{4 m^{2}_e}$. By using the invesrse variable $a^{2} = \frac{1}{b^{2}} = \frac{4 m^{2}_e}{q^{2}}$ one can re-write this 
formula in the form 
\begin{eqnarray}
 W(q) &=& -\frac{4 \pi Q e}{q^{2}} \Biggl\{ 1 + \frac{\alpha}{3 \pi} \Bigl[ -\frac53 + a^{2} + (\frac12 a^{2} - 1) \sqrt{a^{2} + 1} 
 \ln\Bigl(\frac{\sqrt{a^{2} + 1} + a}{\sqrt{a^{2} + 1} - a}\Bigr) \Bigr] \Biggr\} \; \; \label{finalFa} 
\end{eqnarray} 
where $a = \frac{2 m_e}{q}$. The expression $W(q)$ is the Fourier transform of the potential $V(r)$ (Coulomb + Uehling) which is defined by 
Eq.(\ref{first}). The vector-parameter ${\bf q}$ is called the momentum transfer to the heavy nucleus.  

Now, by using Eq.(\ref{finalFa}) we can determine the lowest-order correction on vacuum polarization for a number of QED processes which include 
electron/positron interaction with an external Coulomb field. Indeed, let us assume that we have calculated the cross section $\sigma$ of some 
QED process which includes a direct interaction with an external Coulomb field, or proceeds in the Coulomb field of a heavy (immovable), 
electrically charged center. In this case the corresponding $S-$matrix element includes the factor $\frac{1}{q^{2}} = \frac{1}{\mid {\bf q} 
\mid^{2}}$ which describes interaction between electron and/or positron and an external Coulomb field. The factor $\frac{1}{q^{2}}$ is the Fourier 
transform of the interaction potential. At the next step we want to evaluate the lowest-order vacuum polarization correction to this cross section. 
If such a correction is represented by the Uehling potential, then to acheive our final goal we need to make the following substitution in our 
formula for the cross section and/or for the $S-$matrix element: 
\begin{eqnarray}
 \frac{1}{\mid {\bf q} \mid^{2}} &\Rightarrow& \frac{1}{\mid {\bf q} \mid^{2}} \Biggl\{ 1 + \frac{\alpha}{3 \pi} \Bigl[ -\frac53 + a^{2} + 
 (\frac12 a^{2} - 1) \sqrt{a^{2} + 1} \ln\Bigl(\frac{\sqrt{a^{2} + 1} + a}{\sqrt{a^{2} + 1} - a}\Bigr) \Bigr] \Biggr\} \; \; \label{finalFa1} 
\end{eqnarray} 
In other words, the Fourier transform of the pure Coulomb potential is replaced by the Fourier transform of some `realistic' potential which is the
sum of the Coulomb and Uehling potentials.   

In general, the differential and total cross sections as well as the corresponding $S-$matrix elements of actual QED processes are represented 
in the form of series upon even powers of $\frac{1}{\mid {\bf q} \mid^{2}}$, i.e., for the total cross sections we can write the formula  
\begin{eqnarray}
 \sigma = A_0 + \frac{A_1}{\mid {\bf q} \mid^{2}} + \frac{A_2}{\mid {\bf q} \mid^{4}} + \frac{A_3}{\mid {\bf q} \mid^{6}} + \frac{A_4}{\mid {\bf 
 q} \mid^{8}} + \ldots \; \; \label{sigmaA} 
\end{eqnarray} 
To obtain the lowest order correction on vacuum polarization to this cross section we need to make different substitutions for each term in 
Eq.(\ref{sigmaA}). However, all these substitutions are described by the following general formula   
\begin{eqnarray}
 \frac{1}{\mid {\bf q} \mid^{2 n}} &\Rightarrow& \frac{1}{\mid {\bf q} \mid^{2 n}} \Biggl\{ 1 + \frac{2 n \alpha}{3 \pi} \Bigl[ -\frac53 + a^{2} + 
 (\frac12 a^{2} - 1) \sqrt{a^{2} + 1} \ln\Bigl(\frac{\sqrt{a^{2} + 1} + a}{\sqrt{a^{2} + 1} - a}\Bigr) \Bigr] \Biggr\} \; \; \label{finalFa2} 
\end{eqnarray} 
for $n$ = 1, 2, 3, $\ldots$ and $a = \frac{2 m_e}{q}$. The arising new formula for the total cross section will automatically include the 
lowest-order vacuum polarization correction. A number of useful examples is considered in the next Section. 

\section{Vacuum polarization correction to the Mott scattering formula} 

Let us assume that we have calculated the cross section of some QED process which include interaction of the electron and/or positron with the 
external Coulomb field. Our goal is to evaluate the lowest-order correction on vacuum polarization to this cross section. First, consider 
derivation of the Mott scattering formula for the elastic scattering of the relativistic electron at the immovable Coulomb center. The procedure 
itself is well described in \cite{Grein} (see also \cite{BD}). This allows us to omitt all computational details which are not crucial for our 
current purposes. Below, the notations ${\bf p}_i$ and ${\bf p}_f$ stand for the three-dimensional momenta of the initial and final electron, 
respectively. The differential cross section of the elastic scattering of an electron at external Coulomb potential is \cite{Grein}
\begin{eqnarray}
 \frac{d \sigma}{d \Omega_{f}} &=& \frac{Q^{2} \alpha^{2}}{2 \mid{\bf q}\mid^{4}} \Bigl[ \Bigl( E^{2} + m^{2}_{e} \Bigr) + 
 \frac{\mid{\bf q}\mid^{2}}{4 \sin^{2}\frac{\theta}{2}} \Bigl(1 - 2 \sin^{2}\frac{\theta}{2}\Bigr)\Bigr] \; \; \nonumber \\
 &=& \frac{Q^{2} \alpha^{2}}{2 \mid{\bf q}\mid^{4}} 
 \Bigl( E^{2} + m^{2}_{e} \Bigr) + \frac{Q^{2} \alpha^{2}}{8 \mid{\bf q}\mid^{2}} \Bigl( \frac{1}{\sin^{2}\frac{\theta}{2}} - 2 \Bigr) \; \; 
 \label{Mott}
\end{eqnarray} 
where $E$ is the total energy of the initial and/or final electron, while ${\bf q} = {\bf p}_f - {\bf p}_i$ is the momentum transfered to the 
heavy Coulomb center, e.g., atomic nucleus, and $\theta$ is the scattering angle. In the case of elastic scattering we can write $q^2 = \mid 
{\bf q} \mid^{2} = \mid {\bf p}_f - {\bf p}_i \mid^{2} = \linebreak 2 \mid {\bf p} \mid^{2} \sin^{2}\frac{\theta}{2}$ \cite{Grein}. Now, by 
using the formulas, Eqs.(\ref{finalFa1}) - (\ref{finalFa2}), one finds  
\begin{eqnarray}
 \frac{d \sigma}{d \Omega_{f}} &=& \frac{Q^{2} \alpha^{2}}{2 \mid{\bf q}\mid^{4}} \Bigl( E^{2} + m^{2}_{e} \Bigr) \Biggl\{ 1 + \frac{4 
 \alpha}{3 \pi} \Bigl[ -\frac53 + a^{2} + (\frac12 a^{2} - 1) \sqrt{a^{2} + 1} \ln\Bigl(\frac{\sqrt{a^{2} + 1} + a}{\sqrt{a^{2} + 1} - 
 a}\Bigr) \Bigr] \Biggr\} \nonumber \\
 &+& \frac{Q^{2} \alpha^{2}}{8 \mid{\bf q}\mid^{2}} \Bigl( \frac{1}{\sin^{2}\frac{\theta}{2}} - 2 \Bigr) \Biggl\{ 1 + \frac{2 \alpha}{3 
 \pi} \Bigl[ -\frac53 + a^{2} + (\frac12 a^{2} - 1) \sqrt{a^{2} + 1} \ln\Bigl(\frac{\sqrt{a^{2} + 1} + 1}{\sqrt{a^{2} + 1} - 
 a}\Bigr) \Bigr] \Biggr\} \; \; \; \label{MottCor}
\end{eqnarray} 
where $a = \frac{m_e}{\mid {\bf p} \mid \sin\frac{\theta}{2}}$. This formula contains the lowest-order corection on the vacuum polarization which 
is described by the Uehling potential, Eq.(\ref{U1}). The angular dependence of the differential cross section $\frac{d \sigma}{d \Omega_{f}}$ is 
quite complicated, since the parameter $a$ depends upon $\theta$. Therefore, the differential cross section $\frac{d \sigma}{d \Omega_{f}}$, 
Eq.(\ref{MottCor}), has a slightly different angular dependence (or $\theta-$dependence), than the original formula, Eq.(\ref{Mott}). In principle, 
the actual difference in the cross sections Eq.(\ref{Mott}) and Eq.(\ref{MottCor}) can be observed and measured in modern experiments. 

Analogous formulas for the differential and total cross sections, which explicitly include the lowest-order correction on the vacuum polarization
(or V.P., for short), can be derived for the bremstrahlung, creation and/or annihilation of the $(e^{-}, e^{+})-$pair in the field of a heavy 
atomic nucleus and some other processes. In all these cases we have to make the same substitutions, Eqs.(\ref{finalFa1}) - (\ref{finalFa2}), in 
the original formulas for the cross sections which can be found in the corresponding Chapters from \cite{Grein} (see also \cite{AB} and \cite{BLP}). 
This simple procedure allows one to obtain simultaneously the cross sections of various QED processes themselves and the lowest-order V.P. 
corrections to them. Here we do not to discuss these obvious transformations of the formulas for the cross sections. Instead, we want to show how 
to make the Uehling potential an integral part of the complete and covariant procedure developed for analytical computations of the second-order 
$S-$matrix elements. In other words, we want to include the Uehling potential, Eq.(\ref{U1}), in the set of fundamental and covariant QED rules 
(Feynmann rules) from the very beginning. 

For simplicity, consider the process of pair creation in the field of atomic nucleus (see, e.g., \cite{Grein} - \cite{BLP}). To simplify discussion 
and notations even further, below we shall apply the Feynmann rules in the form presented in \cite{Grein}. First of all, we note that the Uehling 
potential, Eq.(\ref{U1}), is a scalar and static (i.e., time-indepedent) potential. Therefore, the second-order $S-$matrix element in coordinate 
space takes the form 
\begin{eqnarray}
 S_{fi} &=& - e^{2} \sqrt{\frac{4 \pi m^{2}_e}{E_{+} E_{-} \omega V^{3}}} \int \int d^{4}x d^{4}y \Bigl\{ u(p_{-},s_{-}) e^{\imath p_{-} x} 
 \Bigl[(-\imath \varepsilon) \Bigl( e^{-\imath k x} + e^{-\imath k x} \Bigr) \imath S_F(x - y) \; \; \nonumber \\
 & &(-\imath \gamma^{0}) \Bigl( A^{Coul}_{0}(y) + A^{Uehl}_{0}(y) \Bigr) + (-\imath \gamma^{0}) \Bigl( A^{Coul}_{0}(x) + 
 A^{Uehl}_{0}(x) \Bigr) \imath S_F(x - y) \; \; \nonumber \\
 & &(-\imath \varepsilon) \Bigl( e^{-\imath k y} + e^{-\imath k y} \Bigr)\Bigr] v(p_{+},s_{+}) e^{\imath p_{+} y} \Bigr\} \; \; \label{Sfi}
\end{eqnarray} 
where all notations are the same as in \cite{Grein}, $\imath$ is the imaginary unit, while the symbols $A^{Coul}_{0}(x)$ and $A^{Uehl}_{0}(x)$ are 
the scalar component of the Coulomb and Uehling potentials, respectively (other components of these potentials equal zero identically). The rest 
mass of the electron is designated here by $m_e$ (not $m_0$). In our notation the sum of the $A^{Coul}_{0}(x)$ and $A^{Uehl}_{0}(x)$ components 
simply coincides with the three-dimensional potential $V(r)$ defined by Eq.(\ref{first}). By performing the Fourier integration upon the two sets 
of four-dimensional $x$ and $y$ coordinates and by reducing the arising delta-functions one finds the following formula for the second-order 
$S-$matrix element in momentum space
\begin{eqnarray}
 S_{fi} &=& \frac{8 \pi^{2} e^{3}}{\mid {\bf q} \mid^{2}} \delta(E_{+} + E_{-} -\omega) \sqrt{\frac{4 \pi m^{2}_e}{E_{+} E_{-} \omega V^{3}}} 
 \Biggl\{ 1 + \frac{\alpha}{3 \pi} \Bigl[ -\frac53 + a^{2} + (\frac12 a^{2} - 1) \sqrt{a^{2} + 1} \; \; \nonumber \\
 &\times& \ln\Bigl(\frac{\sqrt{a^{2} + 1} + a}{\sqrt{a^{2} + 1} - a}\Bigr) \Bigr] \Biggr\} \; \Biggl\{ u(p_{-},s_{-}) \Bigl[(-\imath 
 \varepsilon) \frac{\imath}{p_{-} - k - m} (-\imath \gamma^{0}) + (-\imath \gamma^{0}) \frac{\imath}{p_{-} + k - m} \; \; \nonumber \\
 &\times& (-\imath \varepsilon)\Bigr] v(p_{+},s_{+}) \Biggr\} \; \; \label{Sfim}
\end{eqnarray}
where ${\bf q} = {\bf p}_{+} + {\bf p}_{-} - {\bf k}$ in this problem and $a = \frac{2 m_e}{q}$. The following steps of the regular procedure 
used to determine the $\mid S_{fi} \mid^{2}$ value and the corresponding cross sections do not change. 

\section{Conclusion} 

We have considered the basic properties of the Uehling potential and derived its different analytical representations. It is shown that the 
Uehling potential can be written in the closed analytical form(s). In fact, there are quite a few of such analytical formulas which differ 
from each other by the relations known for the Bessel and modified Bessel functions. The explicit formulas for the Fourier transform of the 
Uehling potential is also derived in the simple analytical forms, Eqs.(\ref{FUqfinal}) - (\ref{finalFa}). By using these properties of the 
Uehling potentials and formulas for its Fourier transform we have developed the new procedure which allows one to determine simultaneously 
cross sections of a number of QED processes and the lowest-order V.P. corrections to them. Furthermore, the Uehling potential, Eq.(\ref{U1}), 
can directly be included in the set of fundamental and covariant QED rules which is used for analytical calculations of the $S_{fi}$-matrix 
elements at the first stage of the process.


\begin{thebibliography}{99}

\bibitem{Grein} W. Greiner and J. Reinhardt, \textit{Quantum Electrodynamics} (4th ed., Springer Verlag, Berlin, (2009)).

\bibitem{AB} A. Akhiezer and V.B. Berestetskii, \textit{Quantum Electrodynamics} (4th ed., Science, Moscow, (1981)) [in Russian]. 

\bibitem{BLP} V.B. Berestetskii, E.M. Lifshitz and L.P. Pitaevskii, \textit{Relativistic Quantum Theory} (Pergamon Press, Oxford, (1971)). 

\bibitem{Uehl} E.A. Uehling, Phys. Rev. {\bf 48}, 55 (1935). 

\bibitem{Mandl} These units are also called the natural units (see, e.g., F. Mandl and G. Shaw, \textit{Quantum Field Theory} (John Willey 
and Sons Ltd., New York, (1984)). 

\bibitem{Dubl} T. Dubler, K. Kaeser, B. Robert-Tissot, L.A. Schaller, L. Schellenberg and H. Schneuwly, Nucl. Phys. A {\bf 294}, 397 (1978).  

\bibitem{Plum} G. Plunien and G. Soff, Phys. Rev. A {\bf 51}, 1119 (1995).  

\bibitem{Fro3} A.M. Frolov, J. Comput. Science {\bf 5}, 499 (2014). 

\bibitem{Feyn} R.P. Feynman, Phys. Rev. {\bf 76}, 749 (1949); ibid. {\bf 76}, 769 (1949).  

\bibitem{Dyson} F. Dyson, Phys. Rev. {\bf 75}, 1736 (1949).  

\bibitem{AS} \textit{Handbook of Mathematical Functions} (M. Abramowitz and I.A. Stegun (Eds.), Dover, New York, (1972)). 

\bibitem{GR} I.S. Gradstein and I.M. Ryzhik, \textit{Tables of Integrals, Series and Products} (6th revised ed., Academic Press, New York (2000)).

\bibitem{Wats} G.N. Watson, \textit{A Treatise on the Theory of Bessel Functions} (2nd ed., Cambridge at the University Press, Cambridge, UK (1966)). 

\bibitem{Fro1} A.M. Frolov and D.M. Wardlaw. Eur. Phys. Jour. B {\bf 85}, 348 (2012). 

\bibitem{QEDT} In this paper the Fourier transform is defined by Eq.(\ref{Four1}). The inverse Fourier transform is written as a similar multiple 
integral which has the different kernel $\simeq \exp(-\imath {\bf q} {\bf x})$ and the factor $\frac{1}{(2 \pi)^{3}}$ in front of this multiple 
integral. 

\bibitem{Fro2} A.M. Frolov, Can. J. Phys. {\bf 92}, 1094 (2014). 

\bibitem{Pauli} W. Pauli and M. Rose. Phys. Rev. {\bf 49}, 462 (1936). 

\bibitem{BD} J.D. Bjorken and S.D. Drell, \textit{Relativistic Quantum Mechanics} (McGraw-Hill Book Company, New York, (1964)). 

\end{thebibliography}
\end{document}